\documentclass[sigconf]{acmart}
\usepackage{tcolorbox}
\usepackage{fbox}
\usepackage{enumitem}

\usepackage{soul,xcolor}
\setstcolor{red}

\AtBeginDocument{%
  \providecommand\BibTeX{{%
    \normalfont B\kern-0.5em{\scshape i\kern-0.25em b}\kern-0.8em\TeX}}}

\copyrightyear{2024}
\acmYear{2024}
\setcopyright{rightsretained}
\acmConference[ASE '24]{39th IEEE/ACM International Conference on Automated
Software Engineering }{October 27-November 1, 2024}{Sacramento, CA, USA}
\acmBooktitle{39th IEEE/ACM International Conference on Automated Software
Engineering (ASE '24), October 27-November 1, 2024, Sacramento, CA, USA}
\acmDOI{10.1145/3691620.3695065}
\acmISBN{979-8-4007-1248-7/24/10}

\begin{document}

\title{Root Cause Analysis for Microservices based on Causal Inference: How Far Are We?}

\author{Luan Pham}
\orcid{0000-0001-7243-3225}
\affiliation{%
  \institution{RMIT University}
  \city{Melbourne}
  \country{Australia}
}
\email{luan.pham@rmit.edu.au}

\author{Huong Ha}
\orcid{0000-0003-2463-7770}
\affiliation{%
  \institution{RMIT University}
  \city{Melbourne}
  \country{Australia}
}
\email{huong.ha@rmit.edu.au}

\author{Hongyu Zhang}
\orcid{0000-0002-3063-9425}
\affiliation{%
  \institution{Chongqing University}
  \city{Chongqing}
  \country{China}
}
\email{hyzhang@cqu.edu.cn}

\begin{abstract}
Microservice architecture has become a popular architecture adopted by many cloud applications. However, identifying the root cause of a failure in microservice systems is still a challenging and time-consuming task. In recent years, researchers have introduced various causal inference-based root cause analysis methods to assist engineers in identifying the root causes. To gain a better understanding of the current status of causal inference-based root cause analysis techniques for microservice systems, we conduct a comprehensive evaluation of nine causal discovery methods and twenty-one root cause analysis methods. Our evaluation aims to understand both the effectiveness and efficiency of causal inference-based root cause analysis methods, as well as other factors that affect their performance. Our experimental results and analyses indicate that no method stands out in all situations; each method tends to either fall short in effectiveness, efficiency, or shows sensitivity to specific parameters. Notably, the performance of root cause analysis methods on synthetic datasets may not accurately reflect their performance in real systems. Indeed, there is still a large room for further improvement. Furthermore, we also suggest possible future work based on our findings.
\end{abstract}


% \ccsdesc[500]{Software and its engineering~Software performance}


\maketitle

\section{Introduction}
\label{sec:introduction}

\begin{figure}
\includegraphics[width=0.75\columnwidth]{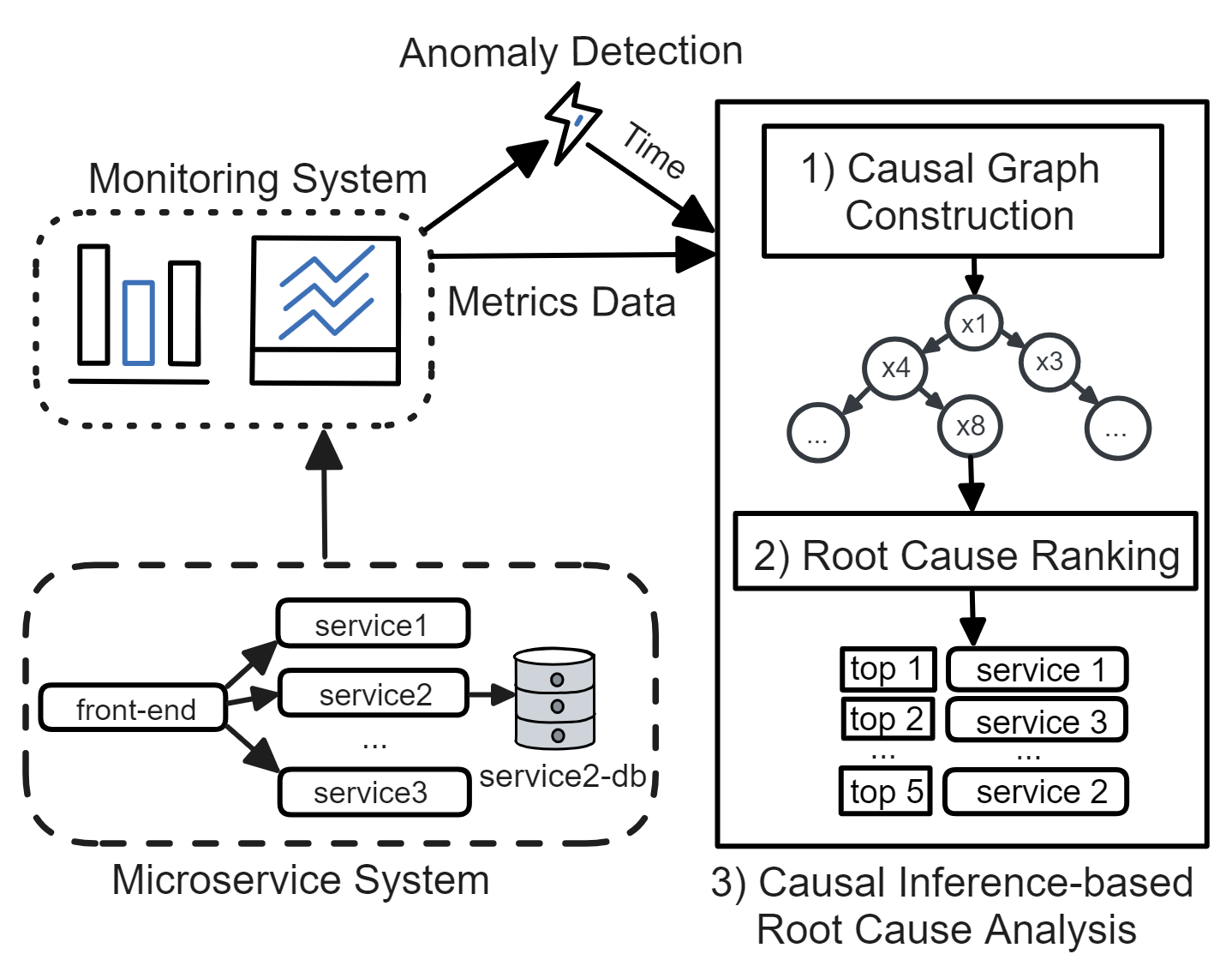}
\vspace{-0.2cm}
\caption{Overview of the causal inference-based root cause analysis for microservice systems using metrics data.} \label{fig:intro}
\vspace{-0.5cm}
\end{figure}

In recent years, microservice architecture has become a popular paradigm in the development of large-scale cloud-based systems (e.g., social networks, online shopping, video streaming services) owing to its scalability, resiliency, and elasticity. A microservice system consists of multiple loosely coupled services where each service can be developed and updated without requiring too much knowledge of the rest of the system. This property makes microservice systems highly adaptable for cloud environments and easy to be deployed, scaled, and maintained by different engineering groups. A large number of prominent enterprises, including Amazon, Netflix, Twitter, and Spotify, have extensively employed a wide range of microservice systems as their core business solutions ~\cite{Soldani2018microservice, Soldani2022rcasurvey}.

Although microservice systems offer various significant benefits, they come with several drawbacks, one of the most notable ones being the challenge of analyzing the root causes of system failures. A typical microservice system can consist of a dozen to hundreds of services, with each service having a large number of metrics to be continuously monitored. Once a failure occurs, it can propagate across the services and affect a large number of metrics, making it especially challenging for engineers to identify the failure's root cause promptly. It was reported that without using an automated tool, it could take engineers at least several hours to identify a failure's root cause~\cite{Azam2022rcd, Wang2018cloudranger}. This delay can affect a large number of users, incurring substantial economic losses and other unintended consequences. It has been reported that a one-hour downtime on Amazon.com could potentially cost up to $100$ million USD~\cite{Chen2020incidentmanagment, Chen2019incienttriage}.

Causal inference-based root cause analysis (RCA) methods for microservice systems via metrics data have attracted increasing attention from researchers in recent years~\cite{Soldani2022rcasurvey, Xin2023CausalRCA, Azam2022rcd, Li2022Circa, Wu2021Microdiag, Ma2019Msrank, Wang2018cloudranger, Meng2020Microcause, run2024aaai, causalai23salesforce}. The main idea is to construct a graph from metrics data to depict the causal relationships among the services and metrics (\textit{causal graph}) and, from this graph, infer the root cause of a failure. A number of methods including CloudRanger~\cite{Wang2018cloudranger}, Microscope~\cite{Jinjin2018Microscope}, MS-Rank~\cite{Ma2019Msrank}, AutoMap~\cite{Ma2020Automap}, MicroCause~\cite{Meng2020Microcause}, and CausalAI~\cite{causalai23salesforce} rely on the Peter-Clark (PC) algorithm~\cite{Spirtes1993Causal} or its variants~\cite{Runge2019PCMCI} to construct the causal graph. Then a scoring method such as random walk~\cite{Spitzer1976randomwalk}, PageRank~\cite{Brin1998Pagerank} or Depth-First Search (DFS) is used to traverse the causal graph to locate the root cause. CIRCA~\cite{Li2022Circa} constructs a causal graph based on domain knowledge and causal assumptions. Then, it uses a regression-based hypothesis testing method to infer the root cause. RCD~\cite{Azam2022rcd} uses the $\Psi$-PC algorithm~\cite{Jaber2020PsiFCI, Spirtes1993Causal} and a divide-and-conquer strategy to infer the failure's root cause. More recently, CausalRCA~\cite{Xin2023CausalRCA} introduces a gradient-based structure learning method to generate a weighted causal graph and combines it with the PageRank algorithm to locate the root cause. Meanwhile, RUN~\cite{run2024aaai} uses neural Granger causal discovery with contrastive learning to construct the causal graph and the PageRank algorithm to infer the root cause.

Despite significant progress, we notice there is a lack of comprehensive evaluation of causal inference-based RCA methods. Existing research works only assess the methods from a limited number of aspects on a restricted set of datasets, and thus, do not provide adequate insights into the capability of these methods. For example, Wu et al.~\cite{Wu2021evalcausal} evaluate six causal inference-based RCA methods. However, it neither evaluates the causal graph construction step nor includes recent proposed methods (e.g., RCD, CIRCA, CausalRCA, RUN), and other important aspects such as the impact of hyperparameter tuning, input data length, among others. Both the work in~\cite{Arya2021evalcausalai} and~\cite{Wang2021evalcausal} only evaluate Granger-based RCA methods for AIOps using time series data obtained via the system log data. \textit{In this work, we aim to understand the current state of causal inference-based RCA methods by thoroughly assessing their performance and the factors that could affect their performance}.

% \st{\textit{In this work, we aim to thoroughly assess the performance of causal inference-based RCA methods to deeply understand how good they are and the factors that could affect their performance.}}

% \red{Q: the authors lack a good problem statement: what is the current state of microservice root cause analysis?}

% \add{A: The purpose of this paper is to understand the current state of causal inference-based RCA methods for microservices. \\ In summary, it is unclear what the current state of RCA methods for microservice systems is. Therefore, in this paper, our goal is to perform a comprehensive evaluation to understand better the current state of causal inference-based RCA methods including their capability in identifying the root causes of various common failures.} \lp{think of how to make this idea more clear}

We conduct a comprehensive evaluation of causal inference-based RCA methods on six synthetic datasets and four datasets from three benchmark microservice systems with different types of failures. First, we assess the performance of various common causal discovery methods in constructing causal graphs for microservice systems from metrics data. Second, we evaluate the performance of existing state-of-the-art causal inference-based RCA methods to understand whether they can accurately locate a failure's root cause. Third, we analyse the runtime of these methods to understand their efficiency. Finally, we investigate different factors that could affect the performance of causal inference-based RCA methods, such as the input data length, the hyperparameter tuning process, and the misspecification of the failure occurrence time. Through extensive experiments, we obtain the following major findings about the causal inference-based RCA methods for microservice systems:

\begin{itemize}[leftmargin=.2in]
    \item All common causal discovery methods have difficulties in working with large graphs and estimating edge directions, suggesting that directly applying causal discovery methods for RCA in large-scale microservices might not be effective. Hyperparameter tuning could help improve the performance of small-scale graphs but fail on large-scale ones.
    \item Some causal inference-based RCA methods are better than others in identifying the root causes of microservice systems' failures but at the cost of %hugely
    losing efficiency or being highly sensitive to some parameters. Large-scale microservice systems remain challenging for causal inference-based RCA methods.
    \item Synthetic datasets used in previous research may not accurately reflect the performance of causal discovery and RCA methods for real-world microservice systems, highlighting the need to reconsider the process of generating synthetic data.
    \item The running time of most causal inference-based RCA methods increases significantly with the size of microservices and the number of metrics. Some methods are always faster than others.
    \item Long input data lengths could significantly improve the performance of causal discovery and causal inference-based RCA methods. Some methods are still effective with shorter data.
\end{itemize}

Based on our study, we identify several challenges of causal inference-based RCA methods in the context of microservice systems and propose possible future research work.
% provide a comprehensive list of the advantages and disadvantages of each causal inference-based RCA method, and the operational personnel can incorporate them into their operational strategies. Furthermore, we also

\noindent In summary, our major contributions are as follows:

\begin{itemize} [leftmargin=.22in]
    \item We conduct a comprehensive evaluation of causal inference-based RCA methods using metrics for microservice systems.
    \item We obtain many important and useful insights from our evaluation and conclude that most existing causal inference-based RCA methods do not have very good performance in all scenarios, especially for large-scale microservice systems.
     \item We suggest future research directions on causal inference-based RCA for microservice systems using metrics data.
\end{itemize}

\section{Causal Inference-Based Root Cause Analysis for Microservice Systems}
\label{sec:background}
\subsection{Problem Statement}

\subsubsection{Key Terminology}

While \textit{failures} represent the actual inability of a service to execute its functions, \textit{faults} correspond to the root causes of such failures (e.g. high CPU utilization, heavy workload, or network congestion)~\cite{Soldani2022rcasurvey, avizienis2004basic}. \textit{Root cause analysis (RCA)} is the process of determining why a failure has occurred~\cite{lee2023eadro}, i.e. finding the root cause of the failure. RCA involves a thorough examination of various monitoring data, i.e. including metrics data. \textit{Metrics} are recorded by the monitoring system and contain various critical information within the microservices systems, such as workload, resource consumption, and response time~\cite{Xin2023CausalRCA}. These metrics are typically represented as multivariate time series, with each time series corresponding to the data collected with a specific metric.

\subsubsection{Problem Formulation}
Considering a large-scale microservice system that consists of $N$ services $\{ s^i \}_{i=1}^N$. At time step $t$, the monitoring system collects $M$ metrics $\textbf{X}^i_t = \{x^{(i, j)}_t\}_{j=1}^{M} (M \geq 1)$ from each service $s^i$. Suppose a failure occurred at time $t_F$. Let us denote the metrics data collected from before the failure, at time step $t_0$, until when the RCA module is invoked at $t_{rca}\ (t_{rca} > t_F)$, as $\textbf{D} = \{ \textbf{X}^1_{t_0}, \dots, \textbf{X}^N_{t_{rca}} \}$. The goal of a metric-based RCA method is to identify the root cause of the failure using the dataset $\textbf{D}$.

The main idea to solve this problem via causal inference is to first construct a causal graph from the dataset \textbf{D} and then use a scoring method to identify the service that is likely to be the root cause of the failure (see Fig. \ref{fig:intro}). In metric-based RCA, \textit{causal graphs} depict cause-and-effect connections, with each node representing a metric of a service and each edge indicating a causal relationship~\cite{Soldani2022rcasurvey}. Causal graphs can capture relationships even among non-communicating services, such as those collocated on the same virtual machine~\cite{Jinjin2018Microscope}. In the sections below, we first describe the existing causal discovery techniques to construct causal graphs from time series data (Sec. \ref{sec:causal-discovery}). Then, we outline the scoring methods to locate the failure's root cause based on the causal graph (Sec. \ref{sec:scoring}), and the state-of-the-art causal inference-based RCA methods for microservice systems (Sec. \ref{sec:causal-rca}).

\subsection{Causal Discovery Methods for Time Series} \label{sec:causal-discovery}

In recent years, causal discovery methods have gained attention for their ability to infer causal relationships from time series data~\cite{Vowels2022causaldiscoverysurvey}. In this section, we briefly summarize representative causal discovery methods commonly used in causal inference-based RCA methods. More information on these causal discovery methods is in our supplementary material (in \href{https://anonymous.4open.science/r/ase24-cfm}{our GitHub page}), Sec. A. 

\textbf{Peter-Clark (PC) Algorithm~\cite{Spirtes1993Causal}.} PC is arguably the most popular causal discovery algorithm. It uses conditional independence testing and a series of rules~\cite{Spirtes1993Causal} to construct the causal graph. PCMCI~\cite{Runge2019PCMCI}, a PC variant, can handle time-lagged causal relations.

\textbf{Fast Causal Inference (FCI) Algorithm~\cite{Spirtes1993Causal}.} Similar to PC, FCI also uses conditional independence testing and a series of rules to construct the causal graph. However, FCI can deal with the presence of confounders, which is an advantage over the PC algorithm. 

\textbf{Granger Algorithm~\cite{Granger1980causal}.} It relies on the concept of \textit{Granger causality}, where a time series causes another if the former provides statistically significant information about future values of the latter. An advanced nonlinear variant, Neural Granger Causal Discovery (NGCD)~\cite{run2024aaai}, can leverage contextual information in temporal data.

\textbf{LiNGAM Algorithm~\cite{Shimizu2006Lingam}.} LiNGAM uses a linear, acyclic structural equation model to construct the causal graph. Similar to PC, it assumes no hidden confounders that affect the time series. 

\textbf{Greedy Equivalence Search (GES) Algorithm~\cite{Chickering2002Ges}.} GES uses a greedy strategy and the Bayesian Information Criterion (BIC)~\cite{Schwarz1978bic} to build the causal graph. Besides, fGES (Fast GES)~\cite{Ramsey2017fges} constructs the causal graph relying on the collider causal structure when orienting edges, making it more computationally efficient.

\textbf{NOTEARS-Low-Rank (NTLR) Algorithm~\cite{fang2023low}.} NTLR is a gradient-based causal discovery method which adapts NOTEARS~\cite{zheng2018dags} with low-rank causal graphs. Note that NLTR has not yet been explored in the RCA literature but we include it to evaluate how a new causal discovery method performs in the RCA task.

\subsection{Scoring Methods for Root Cause Analysis} \label{sec:scoring}

After obtaining a causal graph, a scoring method is used to locate the root cause. We briefly outline below the scoring methods that have been commonly used in causal inference-based RCA.

\textbf{Random Walk.} The main idea of random walk is to walk through all the nodes in the causal graph and randomly choose the next nodes to visit~\cite{Wang2018cloudranger, Jinjin2018Microscope, Meng2020Microcause}. With this strategy, the nodes that are visited most often are considered the root cause of the failure.

\textbf{PageRank.} PageRank assesses the importance of each node in the causal graph based on the number and quality of the incoming edges and identifies nodes with more incoming edges from influential nodes as potential root causes~\cite{Xin2023CausalRCA, Wu2021Microdiag}.

\textbf{Depth First Search (DFS).} DFS traverses all nodes in the causal graph and determines whether they are abnormal via an anomaly detection technique. It then identifies the abnormal sub-graphs, ranks their roots via anomaly scores and determines root causes as the root nodes of the abnormal sub-graphs~\cite{Chen2014Causeinfer}.

\textbf{Hypothesis Testing.} This method formulates a failure as an intervention that alters the pre-failure data distribution. It conducts hypothesis testing to assess if a node's data, after a failure, follows the pre-failure data distribution. Nodes with the most deviation from this distribution are considered root causes~\cite{Li2022Circa, luan2024baro}. 

\subsection{Causal Inference-based Root Cause Analysis Methods for Microservice Systems} \label{sec:causal-rca}

In recent years, many causal inference-based RCA methods have been proposed to analyse and identify the root cause of failures in microservice systems~\cite{Soldani2022rcasurvey, Xin2023CausalRCA, Azam2022rcd, Li2022Circa, Wu2021Microdiag, Ma2019Msrank, Ma2019Msrank, Meng2020Microcause, run2024aaai, causalai23salesforce}. We briefly describe recent causal inference-based RCA methods as follows.

\textbf{PC-based~\cite{Wang2018cloudranger, Jinjin2018Microscope, Meng2020Microcause, Chen2014Causeinfer, Ma2020Automap, Ma2019Msrank}.} These methods use PC or its variant to construct a causal graph from time series metrics data and use a scoring method to locate the root cause. Notable methods include CloudRanger~\cite{Wang2018cloudranger}, 
Microscope~\cite{Jinjin2018Microscope}, CauseInfer~\cite{Chen2014Causeinfer}, AutoMap~\cite{Ma2020Automap}, MicroCause~\cite{Meng2020Microcause} and MS-Rank~\cite{Ma2019Msrank}. 

\textbf{FCI-based~\cite{Chen2019airalert}.} AirAlert proposes to use the FCI algorithm to infer the causal relationships among the metrics data and serves as a diagnosis tool to help engineers identify the root cause easier.

\textbf{Granger-based~\cite{thalheim2017sieve, Wang2021evalcausal, Arya2021evalcausalai, pan2021faster}.} These methods employ the Granger algorithm to derive causal graphs from time series logs data and then identify the root cause using a scoring technique.

\textbf{LiNGAM-based~\cite{Wu2021Microdiag}.} MicroDiag is a popular method following this approach. It uses the DirectLiNGAM algorithm to construct the causal graph from the metrics data and the PageRank method to determine the location of the root cause from the causal graph.

\textbf{MicroCause~\cite{Meng2020Microcause}}. MicroCause uses PCMCI~\cite{Runge2019PCMCI} to construct the causal graph. Then, it applies temporal cause oriented random walk to rank the root causes from the estimated causal graph.

\textbf{CIRCA~\cite{Li2022Circa}.} CIRCA constructs the causal graph from the call graph using operators' knowledge about the system and a mapping of metrics into some defined categories. It formulates a failure as an intervention that alters the metrics data distribution and performs a regression-based hypothesis test to identify the root cause.

\textbf{RCD~\cite{Azam2022rcd}.} RCD follows a divide-and-conquer approach that splits all metrics into smaller chunks and learns a causal graph for each chunk. It employs $\Psi$-PC algorithm~\cite{Jaber2020PsiFCI} to find the root cause within each chunk and then merges all the potential root causes together and runs $\Psi$-PC recursively to identify the final root cause. 

\textbf{CausalRCA~\cite{Xin2023CausalRCA}.} CausalRCA uses a gradient-based variational autoencoder causal structure learning method called DAG-GNN~\cite{Yu2019DagGNN}, to generate the causal graph. To infer the root cause, CausalRCA uses the PageRank algorithm.

\textbf{RUN~\cite{run2024aaai}.} RUN uses NGCD with contrastive learning to construct the causal graph. Then, it applies PageRank with a personalized vector to recommend the top-k root causes.

\textbf{NSigma~\cite{Jinjin2018Microscope}.} NSigma is a hypothesis testing method that uses the z-score to compare the distributions of pre-failure and post-failure data. The higher the z-score, the more likely that metric is the root cause. NSigma does not construct any causal graph.

\textbf{$\epsilon$-Diagnosis~\cite{Shan2019Ediagnosis}.} $\epsilon$-Diagnosis uses a two-sample test algorithm and $\epsilon$-statistics to estimate the similarity between every pair of metrics and rank the root causes based on test scores. Similar to NSigma, $\epsilon$-Diagnosis does not construct causal graphs.

\textbf{BARO~\cite{luan2024baro}.} BARO uses a variant of hypothesis testing technique based on median and interquartile range (IQR), to analyse pre-failure and post-failure data distributions. This makes BARO more resistant to noise compared to NSigma. Similar to NSigma and $\epsilon$-Diagnosis, BARO also does not construct causal graphs.

\textbf{CausalAI~\cite{causalai23salesforce}.} 
CausalAI is an open-source industrial library for causal analysis. To conduct RCA, CausalAI uses PC to build the causal graph using time series metrics data. Subsequently, it derives root nodes from the causal graph as potential root causes.

\section{Study Design}
\label{sec:study-design}
% This section details how we conduct this evaluation study, including dataset preparation, evaluation metrics, and experimental settings.

% \subsection{Research Questions}

% This work aims to comprehensively evaluate causal inference-based RCA methods for microservices to understand their current state and provide useful insights for practitioners and researchers. 
% \hh{The ICSE version has longer research question explanations, should we put some info in the supplementary material?} \lp{I have checked ASE 2023 papers. I see they prefer short version of RQs.}

To understand the current state of causal inference-based RCA methods, we study the following four RQs to thoroughly assess their performance and the factors that could affect their performance:

\begin{itemize}[leftmargin=*]
    \item \textbf{RQ1:} How effective are causal discovery algorithms in constructing causal graphs from time series metrics data? (Sec. \ref{sec:rq1-results})
    
    \item \textbf{RQ2:} How effective are causal inference-based RCA methods in locating the failure's root cause? (Sec. \ref{sec:rca-effectiveness})
    
    \item \textbf{RQ3:} How efficient are causal discovery methods and causal inference-based RCA methods? (Sec. \ref{sec:efficency})
    
    \item \textbf{RQ4:} How do causal discovery and causal inference-based RCA methods perform w.r.t. different input data lengths? (Sec. \ref{sec:eval-input-data})
\end{itemize}

\subsection{Datasets}

\begin{table}[t]
\centering
\caption{Characteristics of synthetic datasets (
\#nodes, \#edges: number of nodes and edges in the graph, \#cases: number of cases in the dataset, \#type: time series data type)
}
\vspace{-0.3cm}
\label{tab:synthetic-data}
\resizebox{0.73\columnwidth}{!}{%
\begin{tabular}{l r r r c}
\hline
\textbf{Name} & \textbf{\#nodes}  & \textbf{\#edges} & \textbf{\#cases}  & \textbf{\#type} \\ \hline \hline
CIRCA10 & 10 & 20  & 200 &  cts \\ \hline
CIRCA50 & 50 & 100  & 200 & cts \\ \hline
RCD10 & 10  & 13-19 & 200 & dct \\ \hline
RCD50 & 50  & 85-104 & 200 & dct \\ \hline
CausIL10  & 10  & 19 & 10 & cts \\ \hline
CausIL50  & 50  & 125 & 10 & cts \\ \hline
\end{tabular}%
}

{\footnotesize (*) 'cts' stands for 'continuous', 'dct' stands for 'discrete'.}
\vspace{-0.2cm}
\end{table}

\begin{table}[t]
\centering
\caption{Characteristics of collected data from benchmark microservice systems (\#metrics, \#svc, \#t\_svc, \#fault: number of metrics, services, targeted services, and fault types).
}
\vspace{-0.3cm}
\label{tab:real-data}
\resizebox{\columnwidth}{!}{%
\setlength\tabcolsep{3pt}
\begin{tabular}{l r r r r r c}
\hline
\textbf{Name} & \textbf{\#metrics}  & \textbf{\#svc}   & \textbf{\#t\_svc}  & \textbf{\#fault} & \textbf{\#cases} & \textbf{\#type}  \\ \hline \hline
Sock Shop 1 & 38 & 13 & 5 & 2 & 50 & cts \\ \hline
Sock Shop 2 & 46 & 15 & 5 & 5 & 125 & cts  \\ \hline
Online Boutique & 49 & 12 & 5  & 5 & 125 & cts \\ \hline
Train Ticket & 212 & 64 & 5 & 5 & 125 & cts \\ \hline
\end{tabular}%
}

{\footnotesize (*) The abbreviation convention is the same as Table \ref{tab:synthetic-data}.}
\vspace{-0.3cm}
\end{table}

\subsubsection{Synthetic Datasets} \label{sec:synthetic-data}

We use three different synthetic data generators from three previous RCA studies~\cite{Li2022Circa, Azam2022rcd, Chakraborty2023CausIL} to create the synthetic datasets: CIRCA, RCD, and CausIL data generators. These data generators are used in various research works to evaluate RCA methods~\cite{Azam2022rcd, Li2022Circa, Chakraborty2023CausIL, liu2023pyrca}. Their mechanisms are as follows:

CIRCA data generator~\cite{Li2022Circa} generates a random causal directed acyclic graph (DAG) based on a given number of nodes and edges. From this DAG, time series data for each node is generated using a vector auto-regression (VAR) model. A fault is injected into a node by altering the noise term in the VAR model for two timestamps. RCD data generator~\cite{Azam2022rcd} uses the pyAgrum package~\cite{pyagrum} to generate a random DAG based on a given number of nodes, subsequently generating discrete time series data for each node, with values ranging from 0 to 5. A fault is introduced into a node by changing its conditional probability distribution. Meanwhile, CausIL data generator~\cite{Chakraborty2023CausIL} generates causal graphs and time series data that simulate the behavior of microservice systems. %It assumes each service has five metrics: workload, CPU usage, memory usage, latency, and error count. 
It first constructs a DAG of services and metrics based on domain knowledge, then generates metric data for each node of the DAG using regressors trained on real metrics data. Unlike the CIRCA and RCD data generators, the CausIL data generator does not have the capability to inject faults.

To create our synthetic datasets, we first generate 10 DAGs whose nodes range from 10 to 50 for each of the synthetic data generators. Next, we generate fault-free datasets using these DAGs with different seedings, resulting in 100 cases for the CIRCA and RCD generators and 10 cases for the CausIL generator. We then create faulty datasets by introducing ten faults into each DAG and generating the corresponding faulty data, yielding 100 cases for the CIRCA and RCD data generators. The fault-free datasets are used to evaluate causal discovery methods, while the faulty datasets are used to assess RCA methods. We use all three dataset generators to alleviate each other's weaknesses, such as diverse causal graph structures and different types of metrics data (continuous and discrete), enabling a more comprehensive assessment of the performance of the studied causal inference-based RCA methods. Table \ref{tab:synthetic-data} shows the characteristics of the synthetic datasets.

% back up for the KDD paper
% We use three different synthetic data generators of three previous RCA studies~\cite{Li2022Circa, Azam2022rcd, Chakraborty2023CausIL} to create synthetic datasets. These data generators are used in various research works to evaluate RCA methods~\cite{Azam2022rcd, Li2022Circa, Chakraborty2023CausIL, liu2023pyrca}. We use all these synthetic datasets to alleviate each other's weaknesses so that we can assess the performance of the studied RCA methods more thoroughly. Each generator produces 10 Directed Acyclic Graphs (DAGs) whose nodes range from 10 to 50. First, we generate fault-free datasets using the ground truth DAGs with different seedings, resulting in 100 cases for CIRCA and RCD data generators and 10 cases for the CausIL data generator. Then, we generate faulty datasets by introducing ten faults into each DAG and generating the corresponding faulty data, yielding 100 cases for CIRCA and RCD data generators. The fault-free datasets are for evaluating causal discovery methods, whilst the faulty ones are for assessing RCA methods. Table \ref{tab:synthetic-data} shows the characteristics of the synthetic datasets. More information about the synthetic datasets is in our supplementary material, Sec. B.1.

\begin{figure}
\includegraphics[width=0.95\columnwidth]{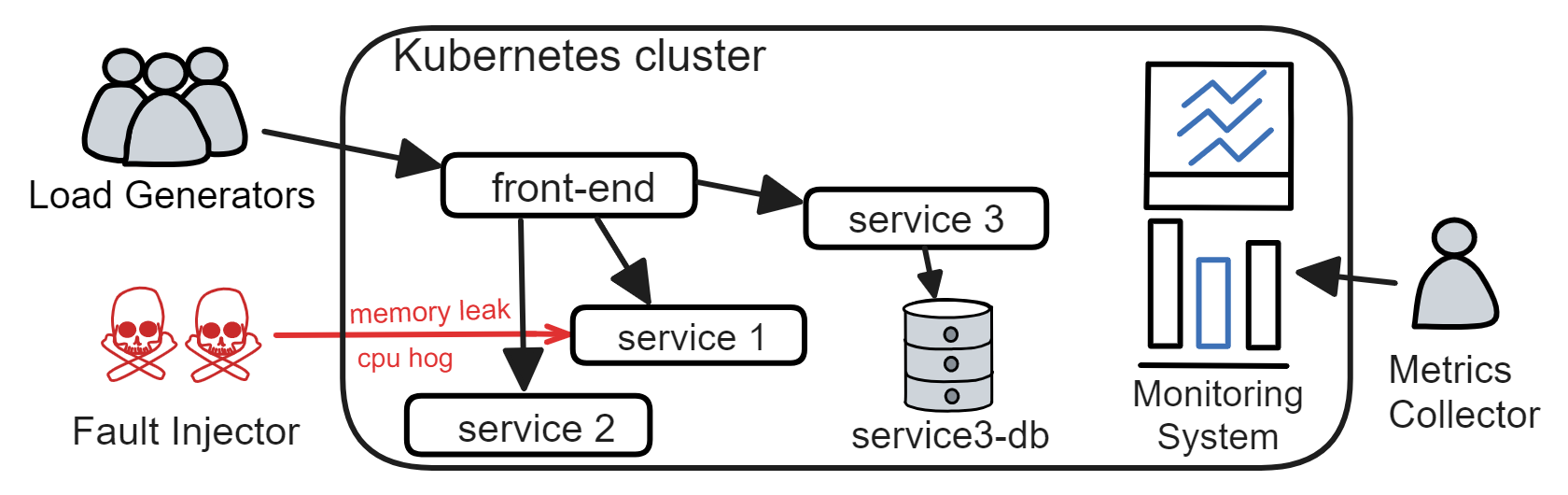}
\vspace{-0.3cm}
\caption{Overview of our setup for microservice systems.} \label{fig:system_setup}
\vspace{-0.4cm}
\end{figure}

\subsubsection{Benchmark Microservice Systems} \label{sec:realworld-data}

We deploy three popular benchmark microservice systems: Sock Shop~\cite{sockshop}, Online Boutique~\cite{ob}, and Train Ticket~\cite{tt}, which are widely used for evaluating RCA methods~\cite{Jinjin2018Microscope, Azam2022rcd, Wu2021Microdiag, Xin2023CausalRCA, wu2022automatic, he2022graph, dan2021practical, yu2021microrank, zhou2018trainticket, Wang2021evalcausal, luan2024baro}. Sock Shop~\cite{sockshop} is a sock-selling e-commerce application that consists of 15 services communicating with each other through HTTP requests. Online Boutique~\cite{ob}, with its 12 services, is an e-commerce application where users can browse items, add them to the cart, and purchase them. Train Ticket~\cite{tt} is one of the largest microservice systems, simulating a train ticket booking system with 64 services. Compared to Sock Shop and Online Boutique systems, Train Ticket system has longer and more complex failure propagation paths. 

To generate metrics data, we first deploy these microservice systems on a four-node Kubernetes cluster hosted by AWS. Next, we use the Istio service mesh~\cite{istio} with Prometheus~\cite{prometheus} and cAdvisor~\cite{cadvisor} to monitor and collect resource-level and service-level metrics of all services, as in previous works~\cite{Azam2022rcd, Xin2023CausalRCA, luan2024baro}. To generate traffic, we use the load generators provided by these systems and customise them to explore all services with 100 to 200 users concurrently. We then introduce five common faults (CPU hog, memory leak, disk IO stress, network delay, and packet loss) into five different services within each system. Finally, we collect metrics data before and after the fault injection operation. An overview of our setup is presented in Fig. \ref{fig:system_setup}. Furthermore, we diversify our datasets by using the available Sock Shop data from a previous study~\cite{Azam2022rcd}, which we refer to as Sock Shop 1. We refer to our Sock Shop data as Sock Shop 2. The statistics of the collected datasets are shown in Table \ref{tab:real-data}.

\subsection{Evaluation Metrics}

\subsubsection{Causal Graph Construction.} We assess the accuracy of the estimated causal graph and its skeleton using F1-score, defined as,
\begin{equation} \nonumber
F1 = \frac{2 \times Pre \times Rec}{Pre + Rec}, \quad Pre = \frac{TP}{TP + FP}, \quad Rec = \frac{TP}{TP + FN},
\end{equation}
where $TP$, $FP$, and $FN$ correspond to true positives, false positives, and false negatives, respectively. $TP$ is the number of correctly identified actual edges, $FP$ is the number of incorrectly identified edges, and $FN$ is the number of missing edges. The F1-score on the skeleton graph, denoted as F1-S, evaluates the accuracy of the edges without considering their directions. The F1-score on the full directed graph, denoted as F1, takes into account the orientations of the edges and penalises for incorrectly estimated directions of correctly identified adjacencies. Following previous works~\cite{Chakraborty2023CausIL, liu2023pyrca}, we also use Structural Hamming Distance (SHD)~\cite{raghu2018evaluation} to assess the estimated causal graph. The SHD score is obtained by summing the missing edges, extra edges, and incorrectly directed edges.

\subsubsection{Root Cause Analysis.} In this work, we evaluate the accuracy in identifying the root cause services as this is a standard practice in related work~\cite{Jinjin2018Microscope, wang2023root, Azam2022rcd, Xin2023CausalRCA, Wang2018cloudranger, Chen2014Causeinfer, Ma2019Msrank, Ma2020Automap}. The root cause service is the service associated with the identified metric from the causal graph~\cite{Azam2022rcd, Ma2020Automap, Xin2023CausalRCA}. Following existing works~\cite{Jinjin2018Microscope, Azam2022rcd, Xin2023CausalRCA, Meng2020Microcause, yu2021microrank, Li2022Circa}, we use two standard metrics, $AC@k$ and $Avg@k$, to assess the performance of the RCA methods. 
$AC@k$ represents the probability the top $k$ results given by a method include the root cause. $AC@k$ scores range from $0$ to 1, and the higher the value, the better the method. Given a set of failure cases A, $AC@k$ is calculated as follows,
\begin{equation} \nonumber
    AC@k = \frac{1}{|A|} \sum\nolimits_{a\in A}\frac{\sum_{i<k}R^a[i]\in V^a_{rc}}{min(k, |V^a_{rc}|)},
\end{equation}
where $R^a[i]$ is the ranking result for the failure case a. $V^a_{rc}$ is the root cause set of case a. 
$Avg@k$, which measures the overall performance of RCA methods, is calculated as $Avg@k = \frac{1}{k}\sum\nolimits_{1\le j\le k} AC@j$.

\vspace{-0.1cm}
\subsection{Experimental Settings}

For PC, FCI, LiNGAM, and CausalRCA methods, we use the implementation published in~\cite{Xin2023CausalRCA}. For Granger, we use the standard implementation from the statsmodels package~\cite{statsmodels}. For PCMCI, we use the implementation in~\cite{Li2022Circa}.  
For fGES and NTLR, we use the source code in~\cite{Chakraborty2023CausIL} and~\cite{zhang2021gcastle}, respectively.
For $\epsilon$-Diagnosis, we use the implementation in~\cite{liu2023pyrca}. 
For RCD, CIRCA, MicroCause, RUN, CausalAI, and BARO, we use their available implementation in~\cite{Azam2022rcd, Li2022Circa, pan2021faster, run2024aaai, causalai23salesforce, luan2024baro}.
Note that for CIRCA, since the call graph is unavailable, thus following~\cite{Azam2022rcd}, we use the PC algorithm to construct this graph. For the hyperparameter settings of the methods, we use the default values suggested by their respective papers. We confirmed the correctness of the source code by reproducing the presented results in the original and related papers. We conduct all experiments on Linux servers equipped with 8 CPUs, 16GB RAM.%, and Ubuntu 22.04.

\section{Results}
\label{sec:results}

%%%%%%%% RQ1. CAUSAL GRAPH CONSTRUCTION PERFORMANCE  %%%%%%%%
\subsection{How Effective are Causal Discovery Algorithms in Constructing Causal Graphs?} \label{sec:rq1-results}

\begin{table*}[]
\centering
\caption{Performance of nine causal discovery methods on synthetic datasets with default settings. Best results are bold. }
\vspace{-0.3cm}
\label{tab:q1-causal-default}
\resizebox{\textwidth}{!}{%
\begin{tabular}{l|rrr|rrr|rrr|rrr|rrr|rrr}
\hline
 & \multicolumn{3}{c|}{CIRCA10} & \multicolumn{3}{c|}{CIRCA50} & \multicolumn{3}{c|}{RCD10} & \multicolumn{3}{c|}{RCD50} & \multicolumn{3}{c|}{CausIL10} & \multicolumn{3}{c}{CausIL50} \\ \hline
\textbf{} & \multicolumn{1}{l}{\textbf{F1}} & \multicolumn{1}{l}{\textbf{F1-S}} & \multicolumn{1}{l|}{\textbf{SHD}} & \multicolumn{1}{l}{\textbf{F1}} & \multicolumn{1}{l}{\textbf{F1-S}} & \multicolumn{1}{l|}{\textbf{SHD}} & \multicolumn{1}{l}{\textbf{F1}} & \multicolumn{1}{l}{\textbf{F1-S}} & \multicolumn{1}{l|}{\textbf{SHD}} & \multicolumn{1}{l}{\textbf{F1}} & \multicolumn{1}{l}{\textbf{F1-S}} & \multicolumn{1}{l|}{\textbf{SHD}} & \multicolumn{1}{l}{\textbf{F1}} & \multicolumn{1}{l}{\textbf{F1-S}} & \multicolumn{1}{l|}{\textbf{SHD}} & \multicolumn{1}{l}{\textbf{F1}} & \multicolumn{1}{l}{\textbf{F1-S}} & \multicolumn{1}{l}{\textbf{SHD}} \\ \hline \hline
PC & \textbf{0.49} & 0.65 & \textbf{16} & \textbf{0.38} & 0.47 & \textbf{104} & 0.3 & \textbf{0.59} & \textbf{14} & 0.24 & \textbf{0.46} & 120 & 0.45 & 0.75 & 16 & 0.3 & 0.46 & 145 \\ \hline
FCI & 0.43 & 0.63 & 19 & 0.33 & \textbf{0.48} & 115 & \textbf{0.36} & \textbf{0.59} & 16 & \textbf{0.3} & \textbf{0.46} & 137 & 0.5 & \textbf{0.76} & 16 & \textbf{0.31} & \textbf{0.47} & \textbf{144} \\ \hline
Granger & 0.46 & 0.6 & 26 & 0.18 & 0.23 & 463 & 0.1 & 0.21 & 19 & 0.19 & 0.42 & \textbf{86} & 0.44 & 0.62 & 28 & 0.13 & 0.22 & 650 \\ \hline
ICALiNGAM & 0.18 & 0.66 & 28 & 0.08 & 0.35 & 283 & 0.19 & 0.46 & \textbf{14} & 0.19 & 0.42 & \textbf{86} & 0.22 & 0.72 & 20 & 0.09 & 0.36 & 281 \\ \hline
DirectLiNGAM & 0.4 & 0.66 & 22 & 0.22 & 0.37 & 249 & 0.19 & 0.45 & \textbf{14} & 0.2 & 0.42 & \textbf{86} & \textbf{0.54} & \textbf{0.76} & \textbf{13} & 0.21 & 0.36 & 263 \\ \hline
GES & 0.42 & 0.66 & 20 & 0.34 & 0.44 & 160 & 0.23 & 0.32 & 15 & 0.23 & 0.32 & 92 & 0.47 & 0.67 & 18 & 0.22 & 0.41 & 205 \\ \hline
fGES & 0.3 & \textbf{0.67} & 22 & 0.18 & 0.44 & 165 & 0.25 & 0.32 & 15 & 0.24 & 0.31 & 92 & 0.36 & \textbf{0.76} & 19 & 0.23 & 0.41 & 195 \\ \hline
PCMCI & 0.12 & 0.18 & 32 & 0.04 & 0.07 & 986 & 0.22 & 0.38 & 44 & 0.06 & 0.11 & 1223 & 0.16 & 0.25 & 35 & 0.06 & 0.11 & 1101 \\ \hline
NTLR & 0.32 & 0.52 & 21 & 0.14 & 0.27 & 131 & 0.19 & 0.34 & 20 & - & - & - & 0.43 & 0.66 & 25 & 0.06 & 0.11 & 570 \\ \hline
\end{tabular}%
}
\vspace{-0.3cm}
\end{table*}

\begin{table*}[]
\centering
\caption{Performance of six causal discovery methods on synthetic datasets with hyperparameter tuning. Best results are bold.}
\vspace{-0.3cm}
\label{tab:q1-causal-tuning}
\resizebox{\textwidth}{!}{%
\begin{tabular}{l|rrr|rrr|rrr|rrr|rrr|rrr}
\hline
 & \multicolumn{3}{c|}{CIRCA10} & \multicolumn{3}{c|}{CIRCA50} & \multicolumn{3}{c|}{RCD10} & \multicolumn{3}{c|}{RCD50} & \multicolumn{3}{c|}{CausIL10} & \multicolumn{3}{c}{CausIL50} \\ \hline
\textbf{} & \multicolumn{1}{l}{\textbf{F1}} & \multicolumn{1}{l}{\textbf{F1-S}} & \multicolumn{1}{l|}{\textbf{SHD}} & \multicolumn{1}{l}{\textbf{F1}} & \multicolumn{1}{l}{\textbf{F1-S}} & \multicolumn{1}{l|}{\textbf{SHD}} & \multicolumn{1}{l}{\textbf{F1}} & \multicolumn{1}{l}{\textbf{F1-S}} & \multicolumn{1}{l|}{\textbf{SHD}} & \multicolumn{1}{l}{\textbf{F1}} & \multicolumn{1}{l}{\textbf{F1-S}} & \multicolumn{1}{l|}{\textbf{SHD}} & \multicolumn{1}{l}{\textbf{F1}} & \multicolumn{1}{l}{\textbf{F1-S}} & \multicolumn{1}{l|}{\textbf{SHD}} & \multicolumn{1}{l}{\textbf{F1}} & \multicolumn{1}{l}{\textbf{F1-S}} & \multicolumn{1}{l}{\textbf{SHD}} \\ \hline \hline
PC & \textbf{0.5} & \textbf{0.66} & \textbf{17} & \textbf{0.31} & 0.36 & 170 & 0.31 & 0.64 & 21 & - & \textbf{-} & - & 0.47 & 0.76 & 17 & 0.22 & 0.38 & 192 \\ \hline
FCI & 0.42 & 0.65 & 20 & 0.28 & 0.43 & \textbf{143} & \textbf{0.49} & \textbf{0.87} & 15 & \textbf{-} & \textbf{-} & - & \textbf{0.53} & \textbf{0.78} & \textbf{16} & \textbf{0.25} & 0.42 & 177 \\ \hline
Granger & 0.38 & 0.58 & 32 & 0.1 & 0.16 & 829 & 0.25 & 0.45 & 30 & 0.07 & 0.14 & 780 & 0.45 & 0.68 & 27 & 0.18 & \textbf{0.44} & \textbf{165} \\ \hline
ICALiNGAM & 0.18 & \textbf{0.66} & 28 & 0.08 & 0.35 & 282 & 0.19 & 0.47 & \textbf{14} & 0.19 & 0.42 & 86 & 0.21 & 0.72 & 20 & 0.1 & 0.37 & 255 \\ \hline
fGES & 0.31 & \textbf{0.66} & 22 & 0.18 & \textbf{0.44} & 164 & 0.31 & 0.39 & 15 & 0.24 & 0.31 & 92 & 0.34 & 0.77 & 19 & 0.23 & 0.41 & 192 \\ \hline
PCMCI & 0.12 & 0.18 & 32 & 0.04 & 0.07 & 925 & 0.22 & 0.38 & 44 & 0.05 & 0.1 & 1225 & 0.16 & 0.25 & 35 & 0.06 & 0.11 & 1119 \\ \hline
\end{tabular}%
}
\\
{\footnotesize \textit{(*)  PC and FCI results on the RCD50 dataset were not obtained due to OOM errors during execution. GES, DirectLiNGAM, and NTLR were excluded for exceeding a 1-hour/case time-out.
}
}
% \vspace{-0.2cm}
\end{table*}

%%%%%%%%%% ANALYSIS %%%%%%%%%%%%%

The performance of causal graph construction directly impacts the performance of causal inference-based RCA methods, yet previous works have overlooked this evaluation \cite{Xin2023CausalRCA, Azam2022rcd, Jinjin2018Microscope, Chen2014Causeinfer, Meng2020Microcause, Wang2018cloudranger}. In this section, we evaluate a comprehensive list of common causal discovery methods: Granger, PC, PCMCI, FCI, DirectLiNGAM, ICALiNGAM, GES, fGES, and NTLR on six synthetic datasets: CIRCA10, RCD10, CausIL10, CIRCA50, RCD50, and CausIL50, to understand their effectiveness. All experiments are repeated 10 times, and we report the average of F1, F1-S, and SHD scores.

In Table \ref{tab:q1-causal-default}, we present the performance of these methods using the default hyperparameters. We have the following findings:

(1) \textbf{PC and FCI yield the best performance.} PC achieves the best score in 7 out of 18 cases, and FCI in 9 cases.

(2) \textbf{All methods struggle with estimating edge directions with Granger, NTLR, LiNGAM/GES-based methods being especially bad at identifying the edge directions.} The F1 scores of all methods are systematically lower than their F1-S scores. This reveals that \textit{we need to consider the capacity of causal discovery methods in estimating the edge directions when developing causal inference-based RCA methods for microservices in future work}.

(3) \textbf{The performance of all methods decrease significantly when the graph size increases.} This highlights that \textit{directly applying common causal discovery methods for RCA may be ineffective in large-scale microservice systems}.

(4) \textbf{There is still ample room for improvement in developing methods to construct a causal graph from metrics data.} The F1 scores of all methods are only within the range from 0.1 to 0.54, which are far from being ideal (the highest possible score is 1).

Furthermore, we also evaluate the studied methods under hyperparameter tuning setting using the Bayesian Information Criterion score~\cite{Schwarz1978bic, biza2020bictuning}, as described in our supplementary material, Sec. C. The experimental results are presented in Table \ref{tab:q1-causal-tuning}. We observe that: 

(5) \textbf{Hyperparameter tuning could improve the construction results of smaller graphs.} The performance of causal discovery methods with tuned hyperparameters on CIRCA10, RCD10, and CausIL10 is mostly better than when using default values.

(6) \textbf{Hyperparameter tuning does not perform well on larger graphs.} In CIRCA50, the tuned PC and FCI drop 23\% and 10\% respectively compared to the default settings. This issue could be due to the complexity of large graphs, which causes challenges for the hyperparameter tuning method to find optimal hyperparameter values. More work needs to be done to improve the effectiveness of these methods for large and complex graphs.

\vspace{-0.2cm}
\begin{tcolorbox}[left=2pt,right=2pt,top=0pt,bottom=0pt]
\textbf{Summary.} 
All common causal discovery methods have difficulties in dealing with large graphs and estimating edge directions. This suggests that directly applying causal discovery methods for 
RCA in large-scale microservice systems may not be effective. Hyperparameter tuning could improve the RCA performance on small graphs but still could not improve the performance on large graphs. {There is still ample room for further research}.
\end{tcolorbox}
\vspace{-0.4cm}

%%%%%%%%%%%%%%%%%% RQ2. RCA PERFORMANCE  %%%%%%%%%%%%%%%%%%

\subsection{How Effective are Causal Inference-based RCA Methods in Locating Root Causes?} \label{sec:rca-effectiveness}

%%%%%%%%%% ANALYSIS %%%%%%%%%%%%%
In this section, we extensively evaluate if state-of-the-art causal inference-based RCA methods can accurately identify the root causes of microservice system failures. The methods we evaluate are: PC-based, FCI-based, Granger-based, LiNGAM-based, fGES-based, NTLR-based, CausalRCA, MicroCause, CIRCA, RCD, NSigma, $\epsilon$-Diagnosis, BARO, CausalAI, and RUN. For the PC / FCI / Granger / fGES / LiNGAM / NTLR-based methods, we use either the popular PageRank or random walk scoring method to locate the root cause from the causal graphs. To assess whether these RCA methods perform better than random, we also include Dummy, a method that randomly chooses a node within the causal graph as the root cause. We thoroughly evaluate these methods using four synthetic datasets (RCD10, RCD50, CIRCA10, CIRCA50) and four datasets collected from three benchmark systems (Online Boutique, Sock Shop 1 \& 2, and Train Ticket), each with various types of failures. All experiments are repeated 10 times, and average results are reported.

Notably, CIRCA, RCD, NSigma, and $\epsilon$-Diagnosis require knowing the exact failure occurrence time $t_F$, which may not be practical as, in practice, this information is often unavailable. Meanwhile, BARO uses a customised anomaly detection technique to estimate the time $t_F$; however, this estimate may not always be accurate. To better understand the performance of these RCA methods, we include the experiments when we misspecify the failure occurrence time $t_F$. We experiment with the following variants of the RCA methods: RCD [$t_\Delta=0$], CIRCA [$t_\Delta=0$], NSigma [$t_\Delta=0$], $\epsilon$-Diagnosis [$t_\Delta=0$], BARO [$t_\Delta=0$] given the perfect $t_F$; and RCD [$t_\Delta=60$], CIRCA [$t_\Delta=60$], NSigma [$t_\Delta=60$], $\epsilon$-Diagnosis [$t_\Delta=60$], BARO [$t_\Delta=60$] given the specified time to be $t_F$ + 60 seconds. The choice of setting $t_\Delta$ to be 60 seconds is inspired by a previous RCA work that uses one-minute sampling intervals~\cite{Li2022Circa}, meaning that the monitoring system collects time series metrics data every 60 seconds.

In Table \ref{tab:q2-rca}, we report the overall performance of all methods in Avg@5. Our major findings are as follows: 

(1) \textbf{CausalRCA, RCD, CIRCA, NSigma, and BARO are generally better than other baselines in identifying root causes of microservice systems' failures.} Meanwhile, PC / FCI / Granger / LiNGAM / fGES / NTLR-PageRank/random walk, CausalAI, RUN, and MicroCause mostly perform similarly to Dummy, meaning they are no better or only slightly better than random selection in identifying the root cause of the datasets and benchmark systems used in our study. To the best of our knowledge, we are the first to use Dummy as a baseline and discover this problem. Examining reported results in previous works \cite{Azam2022rcd, Xin2023CausalRCA, Wu2021Microdiag} yields the same conclusion regarding the effectiveness of the PC / FCI / Granger / LiNGAM / fGES / NTLR-PageRank/random walk approach. \textit{This insight again emphasizes that directly applying these causal discovery methods for RCA in microservice systems may not be effective.}

(2) \textbf{CIRCA and NSigma are sensitive to the specification of the failure occurrence time, especially for large systems like Train Ticket, whilst RCD, $\epsilon$-Diagnosis, and BARO are more robust to this time.} While previous works assume the availability of this information when performing RCA on microservice systems, this finding suggests that \textit{the estimated failure occurrence time might significantly affect the robustness of interventional recognition-based RCA methods}, an aspect that is overlooked in previous works \cite{Azam2022rcd, Li2022Circa}. This finding also implies that \textit{future work should evaluate these RCA methods with different anomaly detectors.}

%%%%%%%%%% TABLE %%%%%%%%%%%%%
\begin{table*}[]
\centering
\caption{Performance of twenty-one RCA methods in terms of Avg@5 on eight datasets.  The fault types CPU, MEM, DISK, DELAY, LOSS, and SIM denote CPU hog, memory leak, disk IO stress, network delay, packet loss, and simulation faults, respectively.}
\vspace{-0.3cm}
\label{tab:q2-rca}
\resizebox{\textwidth}{!}{%
\setlength\tabcolsep{3pt}
\begin{tabular}{l|r|r|r|r|rrrrr|rr|rrrrr|rrrrr}
\hline
 & RCD10 & RCD50 & CIRCA10 & CIRCA50 & \multicolumn{5}{c|}{Online Boutique} & \multicolumn{2}{c|}{Sock Shop 1} & \multicolumn{5}{c|}{Sock Shop 2} & \multicolumn{5}{c}{Train Ticket} \\ \hline
 
 & SIM & SIM & SIM & SIM & \multicolumn{1}{c}{CPU} & \multicolumn{1}{c}{MEM} & \multicolumn{1}{c}{DISK} & \multicolumn{1}{c}{DELAY} & \multicolumn{1}{c|}{LOSS} & \multicolumn{1}{c}{CPU} & \multicolumn{1}{c|}{MEM} & \multicolumn{1}{c}{CPU} & \multicolumn{1}{c}{MEM} & \multicolumn{1}{c}{DISK} & \multicolumn{1}{c}{DELAY} & \multicolumn{1}{c|}{LOSS} & \multicolumn{1}{c}{CPU} & \multicolumn{1}{c}{MEM} & \multicolumn{1}{c}{DISK} & \multicolumn{1}{c}{DELAY} & \multicolumn{1}{c}{LOSS} \\ \hline \hline
Dummy & 0.3 & 0.06 & 0.3 & 0.06 & 0.25 & 0.24 & 0.26 & 0.26 & 0.25 & 0.36 & 0.33 & 0.37 & 0.38 & 0.33 & 0.38 & 0.37 & 0.07 & 0.08 & 0.06 & 0.07 & 0.07 \\
PC-PR & 0.24 & 0.11 & 0.25 & 0.07 & 0.34 & 0.15 & 0.32 & 0.33 & 0.38 & 0.38 & 0.27 & 0.46 & 0.47 & 0.4 & 0.22 & 0.42 & 0.07 & 0.15 & 0.09 & 0.10 & 0.07  \\
PC-RW & 0.13 & 0 & 0.31 & 0.01 & 0.34 & 0.31 & 0.39 & 0.18 & 0.32 & 0.44 & 0.44 & 0.48 & 0.46 & 0.49 & 0.45 & 0.36 & 0.16 & 0.05 & 0.11 & 0.03 & 0.09 \\
FCI-PR & 0.29 & 0.09 & 0.31 & 0.06 & 0.22 & 0.22 & 0.38 & 0.29 & 0.4 & 0.44 & 0.2 & 0.39 & 0.44 & 0.55 & 0.26 & 0.42 & 0 & 0.06 & 0.22 & 0.13 & 0.12 \\
FCI-RW & 0.1 & 0.04 & 0.31 & 0.04 & 0.56 & 0.56 & 	0.56 & 0.36 & 0.36 & 0.32 & 0.32 & 0.48 & 0.48 & 0.48 & 0.48 & 0.48 & 0 & 0 & 0 & 0 & 0 \\
Granger-PR & 0.31 & 0.06 & 0.4 & 0.06 & 0.22 & 0.45 & 0.41 & 0.31 & 0.25 & 0.34 & 0.31 & 0.46 & 0.81 & 0.58 & 0.38 & 0.38 & 0.04 & 0.04 & 0.14 & 0.04 & 0.04 \\
Granger-RW & 0.23 & 0 & 0.3 & 0.01 & 0.34 & 0.31 & 0.34 & 0.14 & 0.2 & 0.56 & 0.54 & 0.48 & 0.46 & 0.49 & 0.45 & 0.36 & 0.16 & 0.05 & 0.11 & 0.02 & 0.13 \\
ICALiNGAM-PR & 0.17 & 0.01 & 0.26 & 0.02 & 0.22 & 0.08 & 0.09 & 0.12 & 0.3 & 0.31 & 0.26 & 0.06 & 0.01 & 0.01 & 0.00 & 0.18 & 0.03 & 0 & 0.16 & 0.03 & 0 \\
ICALiNGAM-RW & 0.13 & 0 & 0.31 & 0.01 & 0.4 & 0.37 & 0.34 & 0.14 & 0.2 & 0.56 & 0.54 & 0.48 & 0.46 & 0.49 & 0.45 & 0.36 & 0.16 & 0.05 & 0.11 & 0.03 & 0.13 \\
fGES-PR & 0.15 & 0.02 & 0.39 & 0.04 & 0.26 & 0.27 & 0.28 & 0.26 & 0.26 & 0.27 & 0.33 & 0.26 & 0.21 & 0.37	& 0.33 & 0.40 & 0.06 & 0.04 & 0.11  & 0.05 & 0.06 \\
fGES-RW & 0.13 & 0 & 0.31 & 0.01 & 0.4 & 0.38 & 0.34 & 0.14 & 0.22 & 0.36 & 0.38 & 0.48 & 0.46 & 0.49 & 0.45 & 0.36 & 0.16 & 0.05 & 	0.04  & 0.03 & 0.13 \\
NTLR-PR & 0.26 & 0.04 & 0.19 & 0.01 & 0.2 & 0.1 & 0.22 & 0.1 & 0.42 & 0.25 & 0.25 & 0.08 & 0.05 & 0.04 & 0.03 & 0.11 & - & - & - & - & - \\
NTLR-RW & 0.14 & 0.05 & 0.3 & 0.02 & 0.34 & 0.31 & 0.34 & 0.14 & 0.2 & 0.56 & 0.54 & 0.48 & 0.46 & 0.49 & 0.45 & 0.36 & - & - & - & - & - \\ 
CausalRCA & 0.1 & 0.04 & 0.05 & 0 & 0.97 & 0.98 & 0.71 & 0.92 & 0.52 & 0.84 & 0.84 & 0.49 & 0.82 & 0.74  & 0.61 & 0.47 & 0.53 & 0.3 & 0.13 & 0.17 & 0.11 \\
CausalAI & 0.08 & 0.06 & 0 & 0 & 0.62 & 0.30 & 0.45 & 0.18 & 0.24 & 0.42 & 0.42 & 0.38 & 0.18 & 0.14 & 0.41 & 0.51 & 0 & 0 & 0.04 & 0.07 & 0.06 \\
RUN & 0.1 & - & 0.3 & - & 0.57 & 0.56 & 0.60 & 0.35 & 0.33 & 0.48 & 0.42 & 0.46 & 0.47 & 0.48 & 0.33 & 0.48 & - & - & - & - & - \\
% RUN & 0.1 & 0 & 0.3 & 0.015 & 0.57 & 0.56 & 0.60 & 0.35 & 0.33 & 0.48 & 0.42 & 0.46 & 0.47 & 0.48 & 0.33 & 0.48 & 0 & 0 & 0 & 0 & 0 \\
MicroCause & 0.13 & 0.04 & 0.37 & 0.09 & 0.34 & 0.45 & 0.37 & 0.57 & 0.42 & 0.4 & 0.55 & 0.47 & 0.43 & 0.33 & 0.22 & 0.46 & - & - & - & - & - \\ \hline 
$\epsilon$-Diagnosis {[}$t_\Delta=60${]} & 0 & 0 & 0 & 0 & 0.23 & 0.03 & 0.12 & 0.16 & 0.16 & 0.5 & 0.66 & 0.51 & 0.38 & 0.49 & 0.46 & 0.46 & 0 & 0 & 0 & 0 & 0 \\
BARO {[}$t_\Delta=60${]} & 0.18 & 0.03 & 0.24 & 0.04 & 0.94 & 0.99 & 0.87 & 0.99 & 0.6 & 0.98 & 0.98 & 0.99 & 0.98 & 0.94  & 1 & 0.88 & 0.81 & 0.99 & 0.77 & 0.82 & 0.72 \\
RCD {[}$t_\Delta=60${]} & 0.48 & 0.21 & 0.18 & 0.03 & 0.74 & 0.59 & 0.67 & 0.67 & 0.5 & 0.43 & 0.67 & 0.43 & 0.35 & 0.63 & 0.5 & 0.44 & 0.11 & 0.09 & 0.1 & 0.12 & 0.19 \\
CIRCA {[}$t_\Delta=60${]} & 0.19 & 0.09 & 0.19 & 0.03 & 0.24 & 0.36 & 0.4 & 0.58 & 0.39 & 0.52 & 0.62 & 0.19	& 0.16 & 0.37 & 0.26 & 0.39 & 0 & 0 & 0.07 & 0.03 & 0.1  \\
NSigma {[}$t_\Delta=60${]} & 0.23 & 0.03 & 0.05 & 0 & 0.16 & 0.24 & 0.43 & 0.55 & 0.38 & 0.15 & 0.33 & 0.27 & 0.22 & 0.23 & 0.37 & 0.5 & 0.03 & 0 & 0.03 & 0.05 & 0.12 \\ \hline
$\epsilon$-Diagnosis {[}$t_\Delta=0${]} & 0 & 0 & 0 & 0 & 0.26 & 0.02 & 0.22 & 0.12 & 0.13 & 0.5 & 0.55 & 0.49 & 0.34 & 0.49 & 0.51 & 0.5 & 0 & 0.02 & 0 & 0 & 0.02 \\
BARO {[}$t_\Delta=0${]} & 0.22 & 0.05 & 0.92 & 0.87 & 0.97 & 1 & 0.91 & 0.98 & 0.67 & 0.99 & 0.99 & 1 & 0.98 & 0.94  & 0.98 & 0.87 & 0.90 & 0.96 & 0.84 & 0.77 & 0.66 \\
RCD {[}$t_\Delta=0${]} & 0.89 & 0.62 & 0.28 & 0.05 & 0.91 & 0.74 & 0.67 & 0.38 & 0.47 & 0.63 & 0.66 & 0.62 & 0.46 & 0.62	& 0.48 & 0.37 & 0.08 & 0.01 & 0.31 & 0.09 & 0.12 \\
CIRCA {[}$t_\Delta=0${]} & 0.34 & 0.08 & 0.34 & 0.06 & 0.94 & 0.97 & 0.7 & 0.92 & 0.55 & 0.7 & 0.8 & 0.97 & 0.98 & 0.92 & 0.98 & 0.88 & 0.66 & 0.93 & 0.64 & 0.64 & 0.57 \\
NSigma {[}$t_\Delta=0${]} & 0.21 & 0.06 & 0.88 & 0.85 & 0.94 & 1 & 0.9 & 0.98 & 0.67 & 0.98 & 0.99 & 0.98 & 0.98 & 0.94 & 0.98 & 0.9 & 0.81 & 0.96 & 0.85 & 0.61 & 0.7 \\ \hline
\end{tabular}%
}
{\footnotesize \textit{(*) 'RW' stands for 'random walk', 'PR' stands for 'PageRank'. (-) MicroCause, RUN, and NTLR-based RCA have missing results because they exceed the limit of 2 hours per case.}} \\ 
\vspace{-0.3cm}
\end{table*}

(3) \textbf{The performance of RCA methods on synthetic datasets may not accurately reflect their performance in real systems.} For example, RCD performs well on its synthetic and other datasets but poorly on CIRCA datasets. NSigma and BARO performs well on CIRCA synthetic datasets and other datasets but poorly on RCD synthetic datasets. Likewise, CausalRCA performs well on Online Boutique, Sock Shop 1 \& 2, but badly on synthetic datasets. These findings suggest \textbf{the generation of synthetic datasets may not be consistent across different works and may not accurately represent metrics data of microservices.} For example, RCD data generator introduces faults by altering the conditional probability of a node whereas real-world anomalies are often characterized by surges in metrics such as high CPU usage, workload drops \cite{lee2023eadro, Soldani2022rcasurvey}. \textit{It is thus important to reconsider the process of generating synthetic data when developing and evaluating RCA methods in future work}.

\begin{tcolorbox}[left=2pt,right=2pt,top=0pt,bottom=0pt]
\textbf{Summary.}
CausalRCA, RCD, CIRCA, NSigma, and BARO are generally the best RCA methods for microservices with metrics data. However, the performance of RCD, CIRCA, and NSigma are sensitive to the estimated failure occurrence time. Future works should evaluate these methods with different anomaly detectors to understand better their effectiveness when using actual estimated failure occurrence time. Existing synthetic datasets used in previous research \cite{Azam2022rcd, Li2022Circa} may not accurately reflect the performance of the RCA methods for real-world microservice systems. Finally, large-scale microservice systems still pose a great challenge for causal inference-based RCA methods.
\end{tcolorbox}
\vspace{-0.2cm}

%%%%%%%%%%%%%%%%%% RQ3. SPEED  %%%%%%%%%%%%%%%%%%
% \vspace{-0.2cm}
\subsection{How Efficient are Causal Discovery and Causal Inference-based RCA Methods?} \label{sec:efficency}

To promptly detect and troubleshoot failures so as to ensure minimal system downtime, it is crucial to develop efficient and effective RCA methods. In this section, we conduct an efficiency analysis of the studied causal discovery and RCA methods using all the available datasets. We record the running time per case of all methods on Linux machines, each with 8 CPU and 16GB memory. 

\subsubsection{Causal Graph Construction}
To evaluate the efficiency of causal discovery methods, we report the runtimes of nine representative methods on six synthetic datasets. Due to space constraints, detailed experimental results are provided in our supplementary material, Table S1. Our findings are:

(1) \textbf{The runtime of all causal discovery methods increases significantly with the graph complexity.} When the size of causal graphs increases from 10 to 50 nodes, these methods become seven to thousands of times slower, taking up to hours to build a causal graph. This observation complements our finding in Section \ref{sec:rq1-results} that \textit{the causal discovery algorithms is neither effective nor efficient on large microservice systems, presenting a challenge of how to select appropriate input metrics for effective RCA in such systems}.

(2) \textbf{The runtime of kernel-based conditional independence testing (PC with KCI \cite{wu2022automatic}, Kernel-based DirectLiNGAM) is prohibitively long.} Running PC with the KCI independence test as described in \cite{wu2022automatic} takes an average of over 1 hour/case to estimate a 10-node graph by CIRCA and RCD synthetic data generators. We can see that \textit{there is a large room for improvement in efficiency if one wants to employ a causal discovery algorithm in their RCA method for diagnosising failures of microservice systems.}

\subsubsection{Root Cause Analysis} \label{sec:speed-rca}
To evaluate the efficiency of RCA methods, in Table \ref{tab:q3-rca-speed}, we provide the running time of twenty-one RCA methods on synthetic and microservices datasets. We found that:

(1) \textbf{NSigma and BARO are consistently faster, followed by RCD whilst CausalRCA, MicroCause, RUN, and NTLR-based methods tend to be the slowest}.  
Based on our observations, NSigma and BARO have a small running time since they do not learn any causal graph, instead, they just compare the normal and abnormal metrics data. RCD also runs fast since it performs hierarchical learning and thus only needs to learn small causal graphs from subsets of metrics data while most other RCA methods learn full causal graphs from all metrics data. CausalRCA, MicroCause, RUN, and NTLR-based methods are significantly slower than others, due to their reliance on advanced causal discovery algorithms like DAG-GNN, PCMCI, NGCD, and NTLR.

(2) \textbf{The running time of most RCA methods increases significantly with the complexity of microservice systems}. Most causal inference-based RCA methods can handle Sock Shop (38-46 metrics) and Online Boutique (49 metrics) within seconds. However, Train Ticket (212 metrics) causes the methods to be much slower; the running time of the methods are from few minutes to one hour.

(3) \textbf{The running time of PC / FCI / Granger / LiNGAM / GES / NTLR-based RCA methods is proportional to the runtime of their corresponding causal discovery methods} (Table S1, supplementary material). Among these RCA methods, the PC-based methods are generally the fastest and NTLR-based methods are the slowest. This is similar to the behaviours of their corresponding causal discovery methods: PC is the fastest and NTLR is the slowest. This shows that \textit{for these RCA methods, their running time depends greatly on the running time of the causal graph construction step}. 

\begin{tcolorbox}[left=2pt,right=2pt,top=0pt,bottom=0pt]
\textbf{Summary.} The running time of causal inference-based RCA methods increases significantly as the size of microservice systems grows. Most causal inference-based RCA methods can handle a small set of metrics (<50) within seconds but slow down remarkably when dealing with larger microservice systems that involve a larger set of metrics. NSigma and BARO are always faster than others whilst CausalRCA, MicroCause, RUN, and NTLR-based methods are usually the slowest.
\end{tcolorbox}

%%%%%%% RQ4. DATA LENGTH %%%%%%%
\subsection{How do Different Methods Perform with Different Input Data Lengths?} \label{sec:eval-input-data}

In this section, we assess whether the performance of causal discovery and causal inference-based RCA methods depend on the input data length. This evaluation is to address the question that \textit{once a failure is detected, whether we can run the RCA methods immediately and thus use a smaller amount of input data or wait to collect more data so as to increase the accuracy of the RCA methods}. For the causal discovery methods, we evaluate their performance with the input data lengths varying from 125 to 4000 data points, corresponding to approximately 2 to 60 minutes of metrics data. For the causal inference-based RCA methods, we evaluate with the input data lengths varying from 60 to 600 data points, corresponding to 1 to 10 minutes of metrics data. It is worth noting that prior works on RCA only evaluate the methods using a smaller number of data points. For example, the works in \cite{wu2022automatic} and \cite{Xin2023CausalRCA} use only 60 data points, while the work in \cite{Azam2022rcd} uses from 535 to 593 data points. All experiments are repeated five times, and we report the average results. Note here we only repeat the experiments 5 times instead of 10 due to the prohibitive running time of this task, i.e., it takes over 400 hours on 8 CPU and 16GB RAM machines for each repeat, making it prohibitive to run with 10 repeats.

\begin{table}[]
\centering
\vspace{0.1cm}
\caption{The running time (in seconds) of twenty-one RCA methods on eight datasets.}
\label{tab:q3-rca-speed}
\vspace{-0.3cm}
\resizebox{\columnwidth}{!}{%
\setlength\tabcolsep{2pt}
\begin{tabular}{l|r|r|r|r|r|r|r|r}
\hline
 & CIRCA10 & RCD10 & CIRCA50 & RCD50 & \multicolumn{1}{c|}{SS1} & \multicolumn{1}{c|}{SS2} & \multicolumn{1}{c|}{OB} & \multicolumn{1}{c}{TT} \\ \hline \hline
PC-PR & 0.18 & 0.05 & 1.78 & 0.39 & 1.53 & 2.16 & 3.39 & 129.65 \\ \hline
PC-RW & 0.17 & 0.05 & 1.77 & 0.39 & 1.63 & 2.24 & 3.53 & 131.27 \\ \hline
FCI-PR & 0.19 & 0.04 & 1.85 & 0.37 & 2.62 & 2.52 & 4.9 & 154.95 \\ \hline
FCI-RW & 0.21 & 0.06 & 1.93 & 0.38 & 2.62 & 2.51 & 5.21 & 152.59 \\ \hline
Granger-PR & 1.05 & 1.3 & 36.32 & 27.55 & 5.72 & 13.28 & 12.9 & 196.3 \\ \hline
Granger-RW & 0.98 & 1.25 & 35.67 & 25.92 & 5.62 & 13.21 & 13.53 & 245.4 \\ \hline
ICA-PR & 0.47 & 0.07 & 6.68 & 6.07 & 0.51 & 2.76 & 1.51 & 16.07 \\ \hline
ICA-RW & 0.44 & 0.07 & 6.11 & 5.60 & 0.53 & 2.77 & 1.55 & 16.39 \\ \hline
fGES-PR & 0.63 & 0.16 & 10.76 & 2.04 & 7.36 & 5.40 & 11.49 & 372.8 \\ \hline
fGES-RW & 0.63 & 0.17 & 10.73 & 2.06 & 7.31 & 5.46 & 11.47 & 381.12 \\ \hline
NTLR-PR & 12.97 & 39.71 & 663.07 & 6179.77 & 487.88 & 471.11 & 448.94 & - \\ \hline
NTLR-RW & 12.7 & 39.26 & 672.11 & 6181.34 &  454.18 & 462.36 & 437.16 & - \\ \hline
CausalRCA & 53.79 & 51.3 & 197.6 & 165.89 & 79.33 & 143.97 & 146.67 & 1326.34 \\ \hline
CausalAI & 0.12 & 0.13 & 6.49 & 1.12 & 7.86 & 17.7 & 16.37 & 643.29 \\ \hline
RUN & 1078.65 & 1095 & - & - & 2051.12 & 4938.75 & 1548.28 & - \\ \hline
MicroCause & 27.6 & 24.69 & 1900.77 & 1430.89 & 75.78 & 206.64 & 257.73 & - \\ \hline
$\epsilon$-Diagnosis & 1 & 1 & 6.3 & 6.62 & 3.54 & 5.42 & 5.28 & 21.92 \\ \hline
RCD & 0.1 & 0.06 & 0.37 & 0.13 & 2.1 & 3.08 & 3.05 & 12.44 \\ \hline
CIRCA & 0.18 & 0.06 & 2.01 & 0.35 & 1.76 & 7 & 4.19 & 3792.29 \\ \hline
NSigma & 0.01 & 0.01 & 0.01 & 0.01 & 0.01 & 0.01 & 0.01 & 0.01 \\ \hline
BARO & 0.01 & 0.01 & 0.01 & 0.01 & 0.01 & 0.01 & 0.01 & 0.01 \\ \hline
\end{tabular}%
}

{\footnotesize \textit{(*) SS1, SS2, OB, TT denote Sock Shop 1, Sock Shop 2, Online Boutique, Train Ticket.}}
\vspace{-0.3cm}
\end{table}

\subsubsection{Graph Construction} 
\begin{figure*}
\includegraphics[width=0.97\textwidth]{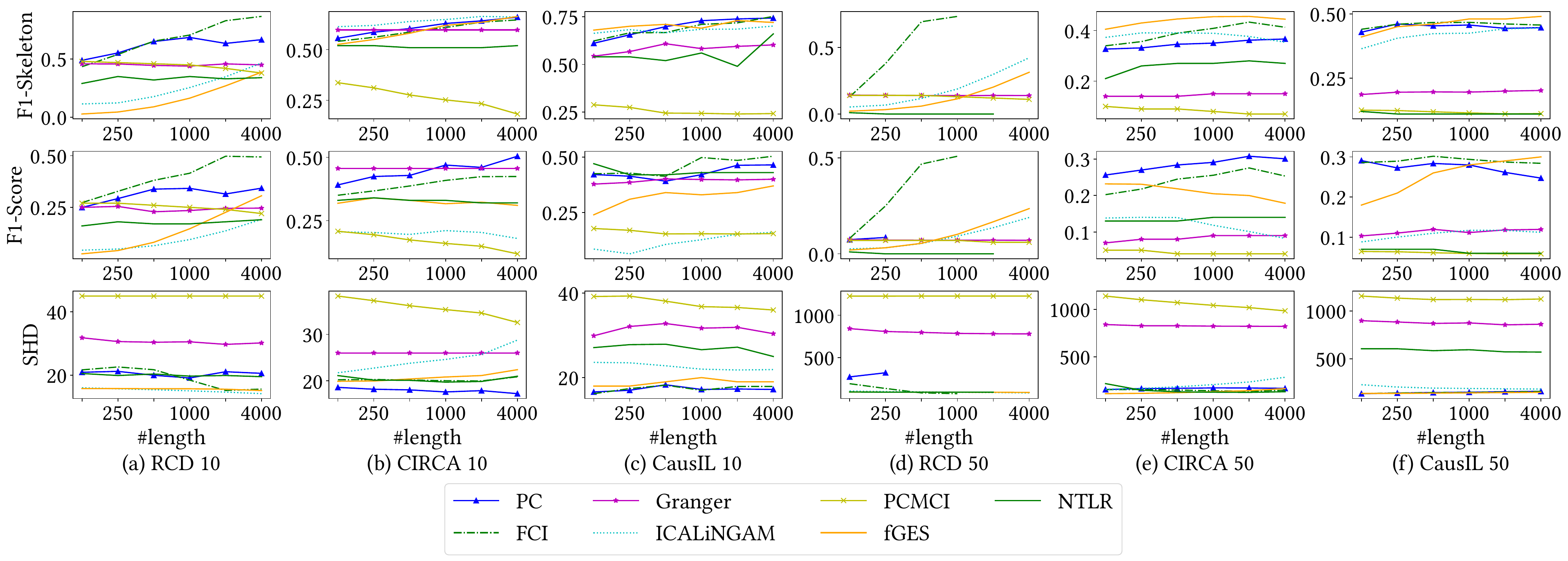}

{\footnotesize \textit{(*) PC, FCI, and NTLR results on RCD50 were partially obtained due to OOM errors during execution and exceeding the time limit.}} 

\caption{Performance of seven causal discovery methods on six synthetic datasets with different data lengths.}  \label{fig:rq4-causal-graph}
\end{figure*}

We plot the performance of seven causal discovery methods (PC, FCI, Granger, ICALiNGAM, PCMCI, fGES, NTLR) on six synthetic datasets. 
The plot is presented in Fig. \ref{fig:rq4-causal-graph}.
%Due to the space limit, we put this plot in our supplementary material, Fig. S2. 
Our observations are:

(1) \textbf{For most datasets, most methods (PC, FCI, ICALiNGAM, fGES) improve their accuracy when being given more input data.} For example, on RCD10, the F1 score of FCI increases from 0.25 to 0.5 when the input data length increases from 125 to 4000.

(2) \textbf{The accuracy of Granger and NTLR appears to plateau with increasing input metrics data whilst PCMCI's accuracy experiences a downward trend.} In all datasets, both Granger and NTLR maintain similar F1, F1-S, and SHD scores across all the input data lengths whilst the F1, F1-S, and SHD of PCMCI often decline with more input data.

(3) \textbf{On CausIL50 dataset, increasing the input data length does not significantly affect the performance of the causal discovery methods.} Most causal discovery methods maintain similar F1, F1-S, and SHD scores when the input metrics data length increases from 125 to 4000.

\begin{figure*}
\includegraphics[width=0.97\textwidth]{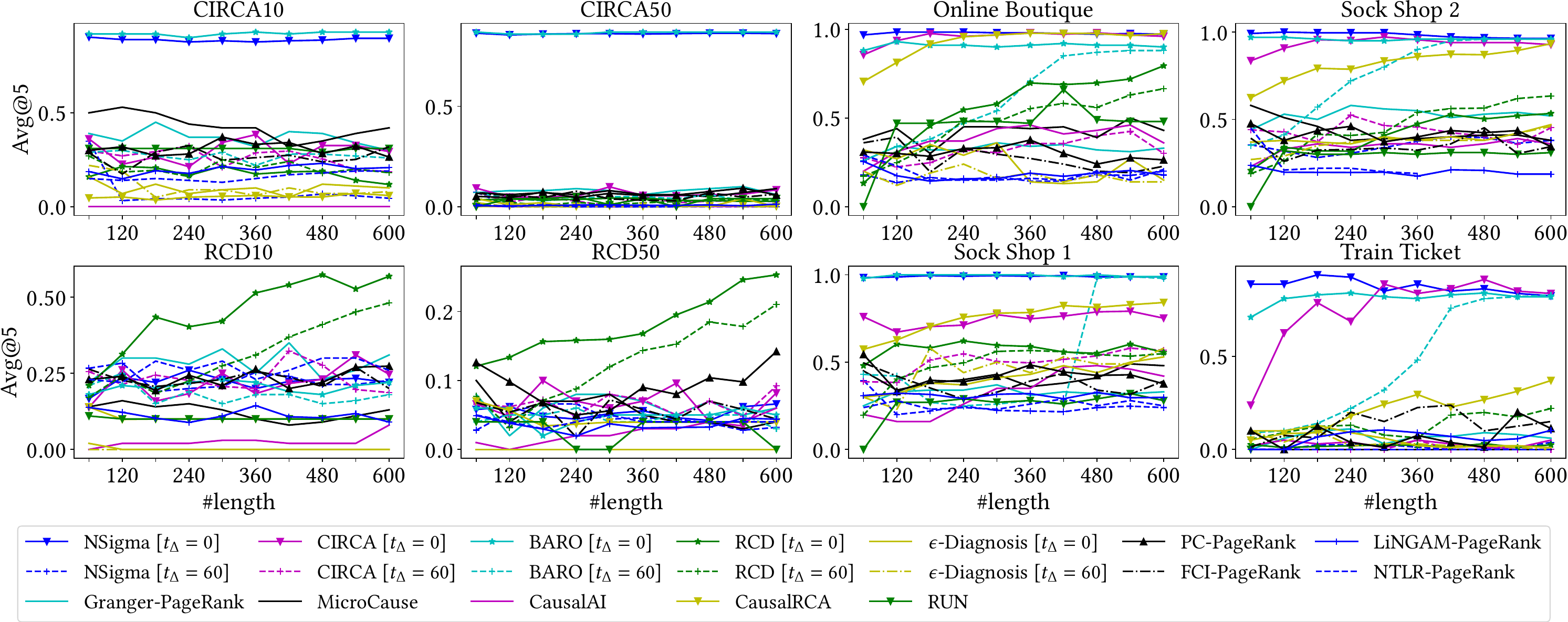}

\vspace{-0.3cm}
\caption{Performance of fourteen RCA methods on eight datasets with different data lengths.}  \label{fig:rq4-rca}
\vspace{-0.3cm}
\end{figure*}

\subsubsection{Root Cause Analysis}
We plot the performance of studied RCA methods on eight different datasets with varying input data lengths in Fig. \ref{fig:rq4-rca}. Our major findings are:

(1) \textbf{With an accurate specification of the failure occurrence time, NSigma, BARO, and CIRCA perform notably well even with less input metrics data.} However, as we emphasized in Section \ref{sec:rca-effectiveness}, their performance drops significantly if there is a misspecification of the failure occurrence time. Notably, increasing the input data lengths, in this case, does not help NSigma and CIRCA, but can help BARO to significantly alleviate the performance degradation. The robustness of BARO in this context can be attributed to its use of median and IQR for hypothesis testing instead of mean and standard deviation as in NSigma and CIRCA.

(2) \textbf{RCD shows remarkable improvements with increased input data lengths.} For example, in Online Boutique, the Avg@5 score of RCD {[}$t_\Delta=0${]} increases from 0.13 to 0.79 when increasing the data length from 60 to 600 data points. This property could be attributed to its usage of $\Psi$-PC \cite{Jaber2020PsiFCI, Spirtes1993Causal}, a novel algorithm that learns a causal graph based on normal data and soft-abnormal data.

(3) \textbf{CausalRCA steadily improves performance when the input data length increases in majority of datasets.} For instance, in Sock Shop 1, its Avg@5 score increases from 0.62 to 0.93 with increased input data length from 120 to 600 data points. One possible reason is that it relies on a deep learning model which typically achieves better performance with more input data.

Based on these findings (1), (2), and (3), we can see that \textit{increasing the data length usually leads to improved performance of RCD, CIRCA, and CausalRCA methods}. However, \textit{this improvement could come with a trade-off that the root cause may not be promptly identified as it generally takes time to collect more abnormal data}. 

(4) \textbf{PC / FCI / Granger / LiNGAM / NTLR-PageRank, RUN, MicroCause, $\epsilon$-Diagnosiss, and CausalAI do not perform well on the studied datasets and adjusting the input data length does not significantly affect their performance.} 
Their performance is consistently low across all datasets and having more metrics data does not seem to help. 

\vspace{-0.1cm}
\begin{tcolorbox}[left=2pt,right=2pt,top=0pt,bottom=0pt]
\textbf{Summary.}
Given an accurate failure occurrence time, NSigma, BARO, and CIRCA perform notably well even with a short metric data lengths, making them ideal candidates for quickly diagnosing failures of microservice systems. Longer input data lengths could significantly improve the performance of both causal discovery and RCA methods. CausalRCA, BARO, CIRCA (given precise failure occurrence time), and RCD exhibit improvements when the input data length increases.
\end{tcolorbox}
\vspace{-0.3cm}

\label{sec:discussion}
\section{Discussion} \label{sec:discussion}
\subsection{The Advantages and Disadvantages of the Studied RCA Methods} \label{sec:advantage-disadvantage} 

Based on our findings, we can conclude that the existing causal inference-based RCA methods for microservice systems need further improvements to work well in practice. We point out the advantages and disadvantages of each method as follows.

\textbf{PC/FCI/Granger/LiNGAM/NTLR-based, RUN, CausalAI, and MicroCause}. These methods learn a complete causal graph from all the metrics data. \textit{With default hyperparameter settings, their performance is not consistent across different datasets; hyperparameter tuning may help to ensure good performance on small-scale microservice systems, but does not help much with large-scale microservice systems}. Finally, for the datasets and microservice benchmark systems used in our studies (10 to 212 nodes), \textit{the performance of these methods is no better or only slightly better than random selection}. This issue might arise from their poor performance in the causal graph construction, as indicated by our findings in Section \ref{sec:rq1-results}. 

\textbf{CIRCA}. CIRCA relies on a user-constructed causal graph, demanding knowledge about microservice structure and metrics, rather than using causal discovery methods. It employs a hypothesis testing scoring method that requires precise failure occurrence time specification for optimal accuracy. However, \textit{CIRCA is highly sensitive to this value; it performs much worse when this value deviates slightly, e.g., 1 minute}. Another drawback of CIRCA is that \textit{it requires the call graph of microservices and manual work in mapping the metrics to construct the causal graph. This leads to difficulties in the adoption of this technique in real-world microservice systems, given their inherent dynamic nature.} Though note that using the PC algorithm to construct the causal graph can still result in a reasonable performance of the algorithm even though it may come with some efficiency trade-offs, particularly in large-scale microservices. 

\textbf{RCD}. The main advantage of RCD is that it does not learn a complete causal graph from all the metrics, but it employs a divide-and-conquer strategy to learn smaller causal graphs from metrics subsets. \textit{This approach makes it highly efficient, especially on large-scale microservices}, i.e., its running time is significantly lower than a majority of RCA methods. \textit{A drawback of RCD is that it requires a sufficient amount of metrics data to perform well.} In practice, this could lead to delays in troubleshooting the failures of microservice systems as it takes time to collect the abnormal/failure data. 

\textbf{NSigma}. NSigma also requires the specification of the failure occurrence time. Similar to CIRCA, its performance is also significantly sensitive to this value. \textit{Given an accurate specification of the failure occurrence time, NSigma can achieve very high accuracy. However, with a slight deviation of the failure occurrence time, e.g., 1 minute, its performance significantly worsens}. Another advantage of NSigma is that \textit{its runtime is very small, making it the fastest method among our studied RCA methods, even on large-scale microservices.}

\textbf{BARO}. BARO can achieve very high accuracy and offer better resistance to the specified failure occurrence time when given enough metrics data. 
This robustness stems from the use of median-based hypothesis testing.
Similar to NSigma, an advantage of BARO is that \textit{its runtime is very small, making it one of the fastest methods among our studied RCA methods.}

\textbf{CausalRCA.} An advantage of CausalRCA is that \textit{it does not require the specification of a failure occurrence time in order to diagnose the root cause.} By using DAG-GNN, a gradient-based causal discovery method, it can diagnose the root cause better than other methods that use PC or FCI. However, due to this characteristic, \textit{CausalRCA is less efficient than other methods.} Its running time is much higher compared to other RCA methods, especially on large-scale microservices with many metrics (>200).

\vspace{-0.2cm}
\subsection{Future Research Directions} \label{sec:future-works}

We identify several challenges of causal inference-based RCA methods in the context of microservices. We outline these challenges in this section and propose possible future research work:

(1) \textbf{Working with Large-scale Microservice Systems.}
Our findings show that most causal discovery methods (to construct the causal graphs) and RCA methods (to identify the failure's root cause) perform well on a small set (<50 metrics) of metrics data and perform badly on a larger set (>200 metrics).
Hence, \textit{selecting an optimal subset of metrics~\cite{thalheim2017sieve}, implementing divide-and-conquer strategies~\cite{Azam2022rcd}, or adding blocked edge sets as domain knowledge~\cite{Chakraborty2023CausIL} may help to reduce the need to construct a full causal graph from all the metrics data}, and thus, improve the performance of root cause identification. Additionally, \textit{using call graphs~\cite{Li2022Circa} or hand-crafted graphs from engineers~\cite{li2022actionable} may also benefit causal inference-based RCA methods}, albeit at a higher cost and risk of error~\cite{li2022actionable,Chen2020incidentmanagment}.

(2) \textbf{Efficiency of the RCA Methods.} Existing works mainly focus on demonstrating the accuracy of the RCA methods without evaluating their efficiency (the running time). Our study suggests that the running time of some RCA methods could be very long, especially when dealing with large-scale microservice systems, and this can lead to a delay in the troubleshooting of the failure. \textit{Future research work could aim to develop RCA techniques that can achieve high accuracy whilst maintaining a reasonable running time.}

(3) \textbf{Sensitiveness to the Failure Occurrence Time.} Our study reveals that \textit{some RCA methods~\cite{Azam2022rcd, Li2022Circa} are sensitive to the failure occurrence time with varying degrees, especially in larger microservice systems}. Future works should extensively evaluate RCA methods within integrated anomaly detection and RCA pipelines. Such evaluations can provide valuable insights into the actual effectiveness of these methods with actual anomaly detectors. 

(4) \textbf{Using a Variety of Datasets from Different Microservice Systems.} Our findings indicate the importance of \textit{using multiple different datasets and microservice benchmark systems to evaluate causal inference-based RCA methods comprehensively}. High performance on one dataset or microservice system does not necessarily reflect high performance on other datasets/systems, as various factors, such as the number of services, metrics, and their inherent dynamic relationships, can significantly affect the RCA performance.

(5) \textbf{Synthetic Dataset Generation}. Since it is not always possible to deploy real microservice systems and simulate a variety of failure scenarios, it is beneficial for the research community to develop methods to generate synthetic data that closely mimics the behaviour of microservice systems. Our results suggest that synthetic datasets used in previous work may not accurately reflect the performance of RCA methods in the real world. \textit{Better methods for generating synthetic data are needed to enhance the development of future causal inference-based RCA methods for microservice systems.}

(6) \textbf{Systematic Hyperparameter Tuning.} In this work, we perform a hyperparameter tuning process via the BIC score~\cite{Schwarz1978bic, biza2020bictuning} and evaluate the performance of common causal discovery methods. Previous works usually choose these hyperparameters empirically, and these chosen values might not be the most optimal choice to achieve the maximal performance. Therefore, \textit{future work can also develop more advanced hyperparameter tuning approaches to improve the performance of causal inference-based RCA methods.}

(7) \textbf{Develop End-to-end Anomaly Detection and RCA}. Future research in RCA for microservices should consider developing more accurate anomaly detection modules. The performance of the RCA pipeline, when considered as a unified system, should be thoroughly evaluated. \textit{This approach would eliminate the need for provided information like the failure occurrence time~\cite{Li2022Circa, Azam2022rcd, luan2024baro}.}

\subsection{Threats to Validity}

We now discuss threats to the validity of our study, along with the means we undertook to mitigate these threats.

\subsubsection{Construct Validity}

The construct validity threat of our evaluation primarily concerns the hyperparameter setting and the evaluation metrics. To address this, we conduct a hyperparameter tuning for studied causal discovery methods. For the studied RCA methods, we use the default values suggested in their papers. We also employ well-established evaluation metrics used in previous works~\cite{Jinjin2018Microscope, Azam2022rcd, Xin2023CausalRCA, Meng2020Microcause, yu2021microrank, Li2022Circa} to compare the performance of causal discovery and RCA methods.

\subsubsection{Internal Validity}

Regarding the studied methods, we re-use the code from various published RCA works, and we have also performed experiments to replicate the results of these source codes to ensure their correctness. To increase the internal validity of our experiment results and avoid the randomness factors in the causal discovery and RCA methods, we repeat the experiments multiple times for each dataset and method and report the average results. We use standard evaluation metrics extensively used in the literature to evaluate the performance of causal graphs and RCA methods. There may be other threats related to the underlying tools, our extracted data, that we have not considered here. To enable exploration of these potential threats and to facilitate replication and extension of our work, we make available our tools and data.

\subsubsection{External Validity}

We evaluate the methods using various synthetic datasets and benchmark systems, along with four common faults. These systems and faults are used in multiple published works on RCA with different characteristics and domains. We acknowledge that different software applications and faults could have different properties and failure propagation mechanisms, which could impact the conclusions in this paper. However, we believe that the datasets and systems we use are representative since they have been used in many previous studies~\cite{Jinjin2018Microscope, Azam2022rcd, Wu2021Microdiag, Xin2023CausalRCA, wu2022automatic, he2022graph, dan2021practical, yu2021microrank, zhou2018trainticket, Wang2021evalcausal} and could help us to derive various important insights for causal inference-based RCA methods for microservice systems.

\subsubsection{Conclusion Validity}

The conclusion validity threat of our evaluation is related to the fault types used in our experiments. Microservice systems can experience different faults that affect the RCA results. To address this, we use five different fault types in our study (CPU, MEM, DISK, DELAY, and LOSS), covering a wide range of different failure scenarios in microservice systems. This allows us to properly evaluate the performance of the studied RCA methods. This is a signifcant improvement over previous works, which typically involved only 2-3 fault types~\cite{Azam2022rcd, Xin2023CausalRCA, Wu2021Microdiag, lee2023eadro}.

\section{Related work}
\label{sec:related-work}
% \section{Related works}

Studying causal inference-based RCA methods and evaluating their performance are important research topics. The work described in~\cite{Soldani2022rcasurvey} conducts a comprehensive survey on anomaly detection and RCA methods for (micro) service-based cloud systems. The studied methods include both causal inference-based and other methods. This survey, however, does not include any evaluation of the RCA methods. 
The works in~\cite{Arya2021evalcausalai} and~\cite{Wang2021evalcausal} evaluate Granger algorithms for AIOps on the Train Ticket microservice system using time series data constructed from logs. Our work, on the other hand, evaluates a wide range of causal discovery and causal inference-based RCA methods on various synthetic datasets and microservice systems. 

% The work in~\cite{Arya2021evalcausalai} evaluates Granger algorithms for AIOps on the Train Ticket system using time series log data. Our work, on the other hand, evaluates a wide range of causal discovery and causal inference-based RCA methods on various synthetic benchmark datasets and microservice systems. The work presented in~\cite{Wang2021evalcausal} also evaluates Granger causal discovery methods on the Train Ticket microservice system with time series data constructed from log data. 

The work described in~\cite{Wu2021evalcausal} evaluates six popular causal discovery methods and combines them with PageRank to identify the root causes from metrics data. These methods are evaluated using the Sock Shop and Train Ticket systems. This work, however, does not evaluate the effectiveness of the constructed causal graph, the scoring methods, recent state-of-the-art causal inference-based methods (RCD, CIRCA, CausalRCA, RUN), and other important aspects such as the impact of hyperparameter tuning, input data length, among others. More recently, there is the work in~\cite{Siebert2023causal} that surveys different causal inference-based methods with the applications in software engineering. This survey, however, does not include any evaluation of these causal inference-based techniques. There are also research works that study and/or evaluate the performance of causal discovery methods for time series data, such as~\cite{Assaad2022causaltimesurvey, Moraffah2021causalsurvey, Glymour2019causal}. However, these works only focus on generic time series data, not metrics data from microservice systems.

\section{Conclusion}
\label{sec:conclusion}

This paper provides a comprehensive evaluation and in-depth analysis of nine causal discovery methods and twenty-one causal inference-based RCA methods for microservice systems using metrics data. We derive many valuable insights from our evaluation and conclude that the performance of existing causal inference-based RCA methods can be further improved to be efficiently and effectively applied in practice. Therefore, more research is needed in this research area. Furthermore, we also release some new datasets to facilitate the development of research in this area. Finally, we suggest some possible future research directions. We believe our contributions advance the understanding of causal inference-based RCA methods for microservices and pave the way for more robust and impactful solutions. 

\section{Data Availability}
We have open-sourced our evaluation framework, RCAEval, which is available on GitHub \cite{rcaevalgithub}. Additionally, an immutable artifact is available on Zenodo \cite{rcaevalgithubzenodo}, together with our experimental datasets~\cite{rcaevaldatasets}.

\begin{acks}
This research was supported by the Australian Research Council Discovery Project (DP220103044), AWS Cloud Credit for Research. We also thank anonymous reviewers for their insightful and constructive comments, which significantly improve this paper. 
\end{acks}

\bibliographystyle{ACM-Reference-Format}
\bibliography{reference}

\end{document}